\documentclass[12pt,preprint]{aastex}

\shorttitle{DM--DE coupling biasing parameter estimation}
\shortauthors{La Vacca et al.}

\begin{document}

\title{Dark Matter--Dark Energy coupling biasing parameter estimates from CMB
data}

\author{Giuseppe La Vacca\altaffilmark{1}}
\affil{Theoretical and Nuclear Physics Department, Pavia University,
Via A. Bassi 6, 20700 Pavia, Italy}
\email{giuseppe.lavacca@mib.infn.it}

\author{Loris P.L. Colombo \altaffilmark{2}}
\affil{Department of Physics \& Astronomy,
University of Southern California, Los Angeles, CA 90089-0484}
\email{colombo@usc.edu}

\author{Luca Vergani}
\affil{Physics Dep.~G.~Occhialini, Milano--Bicocca University, Piazza
della Scienza 3, 20126 Milano, Italy}
\email{luca.vergani@mib.infn.it}

\and

\author{Silvio A. Bonometto \altaffilmark{1}}
\affil{Physics Dep.~G.~Occhialini, Milano--Bicocca University, Piazza
della Scienza 3, 20126 Milano, Italy}
\email{silvio.bonometto@mib.infn.it}

\altaffiltext{1}{I.N.F.N., sez. Milano--Bicocca, Piazza della Scienza 3,
20126 Milano, Italy}
\altaffiltext{1}{Physics Dep.~G.~Occhialini, Milano--Bicocca University, Piazza
della Scienza 3, 20126 Milano, Italy}

\begin{abstract}
When CMB data are used to derive cosmological parameters, their very
choice does matter: some parameter values can be biased if the
parameter space does not cover the ``true'' model. This is a problem,
because of the difficulty to parametrize Dark Energy (DE) physics. We
test this risk through numerical experiments. We create artificial
data for dynamical or coupled DE models and then use MCMC techniques
to recover model parameters, by assuming a constant DE state parameter
$w$ and no DM--DE coupling. For the DE potential considered, no
serious bias arises when coupling is absent. On the contrary,
$\omega_{o,c}$, and thence $H_o$ and $\Omega_{o,m}$, suffer a serious
bias when the ``true'' cosmology includes even just a mild DM--DE
coupling. Until the dark components keep an unknown nature, therefore,
it can be important to allow for a degree of freedom accounting for
DM--DE coupling, even more than increasing the number of parameters
accounting for the $w(a) $ behavior.
\end{abstract}


\section{Introduction}
Scarce doubts remain that Dark Energy (DE) exists. Not only SNIa data
indicate an accelerated cosmic expansion (Perlmutter et al. 1997,
1998, Riess et al. 1998, Foley et al. 2007); also CMB and deep sample
data show a clear discrepancy between the total density parameter
$\Omega_o,$ approaching unity, and the matter density parameter
$\Omega_{o,m} \sim 0.25$--$0.3~$ (see, {\it e.g.}, Spergel et
al. 2007). DE covers this gap; its state parameter $w\equiv
p_{de}/\rho_{de}$ must approach $-1$ today, so apparently excluding
that DE is made of free particles ($p_{o,de}, ~\rho_{o,de}$: DE
pressure, energy density). The true nature of DE is however still
elusive; a false vacuum and a self--interacting scalar field are among
the most popular hypotheses for it (Wetterich 1988, Ratra \& Peebles
1988).

In this paper we explore some possible consequences of our poor
knowledge of DE nature. In particular we test the risk that other
cosmological parameter estimates are biased by a inadequate
parametrization of the DE component. We shall see that this risk is
real.

Theoretical predictions had an astonishing success in fitting CMB
data. For instance, the SW effect, predicting low--$l$ $C_l$ data, or
primeval compression waves, predicting $C_l$ peaks and deeps, were
clearly detected. There is little doubt that we are exploring the
right range of models.

When we investigate DE nature through CMB data, we must bear in mind
that they were mostly fixed at a redshift $z \sim 1100\, ,$ when the
very DE density should be negligible, and so affects peak and deep
positions indirectly, through the values of $\omega_{o,c}\equiv
\Omega_{o,c} h^2$ and $\omega_{o,b}\equiv \Omega_{o,b} h^2$ (here $
H_o = 100\, h\, $km/s/Mpc is the present Hubble parameter;
$\Omega_{o},~ \Omega_{o,c},~ \Omega_{o,b}$ are the present total, CDM,
baryon density parameters).  Later information on DE state equation,
conveyed by the ISW effect, is seriously affected by cosmic variance
and often relies on the assumption that a single opacity parameter
$\tau$ can account for reionization, assumed to be (almost)
instantaneous.  Accordingly, if we assume dynamical DE (DDE), due to a
scalar field $\phi$ self interacting through a potential $V(\phi),$
CMB data allow to exclude some interaction shape, {\it e.g.}
Ratra--Peebles (1988) potentials with significantly large $\Lambda$
energy scales, but hardly convey much information on potential
parameters.

In spite of that, when we choose a DE potential or a specific scale
dependence of the DE state parameter $w(a),$ we risk to bias the
values of other cosmological parameters, sometimes leading to
premature physical conclusions. An example is the value of the
primeval spectral index for scalar fluctuation $n_s$. Using WMAP3 data
(Spergel et al. 2007) and assuming a $\Lambda$CDM cosmology, the value
$n_s=1$ is ``excluded'' at the 2--$\sigma$ confidence level. On the
contrary, Colombo \& Gervasi (2007) showed that this is no longer true
in a DDE model based on a SUGRA potential (Brax \& Martin 1999, 2001;
Brax, Martin \& Riazuelo 2000), whose likelihood was the same of
$\Lambda$CDM.

The risk that our poor knowledge of DE nature biases parameter
determination is even more serious if DM--DE coupling is allowed.
Coupled DE (CDE) cosmologies were studied by various authors (see,
{\it e.g.}, Wetterich 1995, Amendola 2000, Bento, Bertolami \& Sen
2002, Macci\`o et al. 2004).

While DDE was introduced in the attempt to ease DE {\it fine--tuning}
problems, CDE tries to ease the {\it coincidence} problem. Let us then
parametrise the strength of DM--DE coupling through a parameter
$\beta,$ defined below. When $\beta$ is large enough, DM and DE scale
(quasi) in parallel since a fairly high redshift. In turn this
modifies the rate of cosmic expansion whenever DM and/or DE
contributions to the total energy density are non--negligible, so that
limits on $\beta$ can be set through data.

The range allowed ($\beta < 0.10$--0.12; Mainini, Colombo \& Bonometto
2005, Mainini \& Bonometto 2007, see also Majerotto, Sapone \&
Amendola 2004, Amendola, Campos \& Rosenfeld, 2006), unfortunately, is
so limited that DM and DE are doomed to scale differently, but in a
short redshift interval. Clearly, this spoils the initial motivation
of coupling, but, once the genie is outside the lamp, it is hard to
put him back inside: even though the coupling solves little conceptual
problems, we should verify that no bias arises on the other parameters,
for the neglect of $\beta$'s consistent with data. This is far from
being just a theoretical loophole, the still unknown physics of the
dark components could really imply the presence of a mild DM--DE
coupling, and its discovery could mark a step forwards in the
understanding of their nature.

Here we test this possibility by performing some numerical
experiments. We assume DDE and CDE due to a SUGRA potential and use
MCMC techniques to fit the following parameter set: $\omega_{o,c} $,
$\omega_{o,b} $, $\tau$, $\theta$, $n_s$, $A_s$ and (constant) $w$;
$\theta$ is the angular size of the sound horizon at recombination
(see however below), $n_s$ and $A_s$ are spectral index and amplitude
of scalar waves, no tensor mode is considered.

The plan of the paper is as follows: In Section 2 we discuss how
artificial data are built, outlining the models selected, the DDE
potential used and the sensitivity assumed. In Section 3 we briefly
debate the features of the MCMC algorithm used and illustrate a test
on its efficiency, also outlining the physical reasons why some
variables are more or less efficiently recovered. In Section 4 we
discuss the results of an analysis of DDE artificial data, against the
$w = {\rm const.}$ assumption. In Section 5 we briefly summarize why
and how CDE models are built and do the same of Sec.~4 for CDE
models. This section yields the most significant results of this
work. In Section 6 we draw our conclusions.

\section{Building artificial data}
In order to produce artificial data we use CAMB, or a suitable
extension of it (see below), to derive the angular spectra
$C_l^{(TT)}$, $C_l^{(EE)}$, $C_l^{(TE)}$ for various sets of parameter
values. Artificial data are then worked out from spectra, according to
Perotto et al. (2006)$\, $.

The analysis we report is however mostly based on a single choice of
parameters, WMAP5 inspired (see Komatsu et al., 2008, Spergel et
al. 2007), in association with three different $\beta$'s. Two other
parameter choices will also be used: (i) to test the efficiency of the
MCMC algorithm; (ii) to add results for a still lower $\beta$ value,
so strengthening our conclusions.

The parameter sets for the most general case and the (ii) case are
shown in Table~1. In these cases DE is due to a scalar field
self--interacting through a SUGRA potential
\begin{equation}
V(\phi) = (\Lambda^{\alpha+4}/\phi^\alpha) \exp(4\pi\, \phi^2/m_p^2)~,
\label{sugra}
\end{equation}
($m_p:$ Planck mass) which has been shown to fit CMB data at least as
well as $\Lambda$CDM (Colombo \& Gervasi 2007); as outlined in the
Table, we input the value of the energy scale $\Lambda;$ the
corresponding $\alpha$ value is determined by the program itself; the
choice $\Lambda = 1\, $GeV is consistent with Colombo \& Gervasi
(2007) findings and is however scarcely constrained by data. The
meaning of the coupling constant $\beta$ is discussed at the beginning
of Section 5~.

\begin{table}[!hbt]
\begin{center}
\begin{tabular}{ccc}
\hline \hline
& Model A & Model B \\
\hline
$\Omega_o$ & 1 & 1 \\
$10^2 \, \omega_{o,b}$ & 2.273 & 2.4 \\
$\omega_{o,c}$ & 0.1099 & 0.11 \\
$100 \, h$ & 71.9 & 85 \\
$\tau_{opt}$ & 0.087 & 0.15 \\
$\ln(10^{10} \, A_s)$ & 3.1634 & 3.1355  \\
$n_s$ & 0.963 & 1 \\
$\Lambda /$GeV & 1 & 1 \\
\hline
 & 0\, \,   &  \\
$\beta$& 0.05 &  0.02 \\
& 0.10 &    \\
\hline\hline
\end{tabular}
\caption{\textsl{Cosmological parameters for artificial CMB data.}}
\label{tabin}
\end{center}
\vskip -.8truecm
\end{table}

Starting from the spectral components and assuming that cosmic
fluctuations are distributed according to a Gaussian process, we generate {\it
realizations} of the coefficients of the spherical harmonic expansions for the
temperature and E--polarization fields, according to the expressions
\begin{equation}
a_{lm}^{(T)} = \sqrt{C_l^{(TT)}} g_{lm}^{(1)}~,
\label{almt}
\end{equation}
\begin{equation}
a_{lm}^{(E)} = \left[C_l^{(TE)}/C_l^{(TT)}\right] \sqrt{C_l^{(TT)}}
g_{lm}^{(1)} + \sqrt{C_l^{(EE)} -
\left[\left(C_l^{(TE)}\right)^2/C_l^{(TT)} \right] } g_{lm}^{(2)}~,
\label{alme}
\end{equation}
where both $g_{lm}^{(i)},$ for any $l$ and $m$, are casual variables,
distributed in a Gaussian way with null averages and unit variance, so
that the equality $\langle g_{lm}^{(i)} g_{lm}^{(j)} \rangle =
\delta^{ij}$ is approached when the number of realizations increases.
Together with eqs.~(\ref{almt}) and (\ref{alme}) this grants that
$$
\langle a_{lm}^{(T)} a_{l'm'}^{(T)*} \rangle = C_l^{(TT)} \delta_{ll'}
 \delta_{mm'}~,~~
\langle a_{lm}^{(E)} a_{l'm'}^{(E)*} \rangle = C_l^{(EE)} \delta_{ll'}
 \delta_{mm'}~,~~
$$
\begin{equation}
\langle a_{lm}^{(T)} a_{l'm'}^{(E)*} \rangle = C_l^{(TE)} \delta_{ll'}
 \delta_{mm'}~,
\label{aver}
\end{equation}
if averages are taken over an ``infinite'' set of sky realizations.
>From a single realization, we can define estimators of the power as
$$
(2l+1) \hat C_l^{(TT)} = \sum_m a_{lm}^{(T)} a_{lm}^{(T)*}  ~,~~
(2l+1) \hat C_l^{(EE)} = \sum_m a_{lm}^{(E)} a_{lm}^{(E)*} ~,~~
$$
\begin{equation}
(2l+1) \hat C_l^{(TE)} = \sum_m a_{lm}^{(T)} a_{lm}^{(E)*} ~.
\label{sum}
\end{equation}
Taken independently from each other, $ \hat C_l^{(TT)}$ and $ \hat
C_l^{(EE)}$ at a given $l$ are the sum of the squares of $2l +1$
Gaussian random variables, so they are distributed according to a
$\chi^2$ with $2l +1$, which approaches a Gaussian distribution
centered around the fiducial $ C_l^{(TT)}$, $ C_l^{(EE)}$, values, as
$l$ increases.  If we consider $T$ and $E$ simultaneously, the set of
estimators $ \hat C_l^{(TT)}$, $ \hat C_l^{(EE)}$, $ \hat C_l^{(TE)}$
follow a Wishart distribution (see, e.g., Percival \& Brown 2006).

We consider here an idealized full--sky experiment characterized by:
(i) finite resolution, that we shall set through a Full Width Half
Maximum angle $\vartheta_{FWHM}$, where the antenna sensitivity is
reduced to $50\, \%,$ assuming a circularly symmetric Gaussian beam
profile; (ii) background noise due to the apparatus, that we shall
assume to be {\it white}, {\it i.e.} $l$--independent with an assigned
variance~$\sigma^2_N.$

More in detail: for what concerns the Gaussian and circularly
symmetric beam profile, its window function reads
\begin{equation}
_s B_l = e^{-[l(l+1) - s^2] \sigma^2/2}
~,~~{\rm with} ~~~~
\sigma \equiv \vartheta_{FWHM}/\sqrt{8\, \ln 2}~,
\end{equation}
$s$ being the spin of the signal (0 for anisotropy or
$E$--polarization spectra; its value however matters just for the
lowest $l$'s). For what concerns noise, we shall consider the
coefficients $n_{lm}$ such that
\begin{equation}
\langle {a_{lm}^{(R)\, }} n_{l'm'}^{(S)*} \rangle = 0
~,~~~~
\langle {n_{lm}^{(R)\, }} n_{l'm'}^{(S)*} \rangle = N_l^{(RS)} \delta_{ll'}
\delta_{mm'}~,
\end{equation}
with $N_l^{(RS)} = \sigma_N^2 \delta^{RS}$, and $R$ and $S$ stand for
either $T$ or $E$. As the sum of two independent Gaussian random
variables is still Gaussian distributed, the statistics of the
(beam--convolved) CMB +white noise field are given by ${\cal
  C}_l^{(RS)} = C_l^{(RS)}|_0 B_l |^{2} + N_l^{(RS)}$. In an analogous
manner from $\hat C_l^{(RS)}$ we can define $\hat {\cal
  C}_l^{(RS)}$. In the following we will consider both {\it fiducial},
i.e. ${\cal C}_l^{(RS)}$, or {\it realized}, $\hat {\cal C}_l^{(RS)}$, 
model data sets.

Under these simplified assumptions, the characteristics of an experiment
are completely defined by the values of $\vartheta_{FWHM}$ and
$\sigma_N^2$.  Here we shall take $\vartheta_{FWHM} \simeq 7.0',$
while $\sigma_T^2 \simeq 3 \times 10^{-4} (\mu {\rm K})^2$ and
$\sigma_P^2 \simeq 6 \times 10^{-4} (\mu {\rm K})^2$. These can be
considered conservative estimates of sensitivity in the forthcoming
Planck experiment.

\section{The MCMC algorithm}
When the number $\cal N$ of the parameters to be determined from a
given data set is large, the whole $\cal N$--dimensional parameter
space cannot be fully explored within a reasonable computing time.
MonteCarlo Markov Chain (MCMC) algorithms are then used, whose
efficiency and reliability have been widely tested.

In this work we used the MCMC engine and statistical tools provided by
the CosmoMC package (Lewis \& Bridle 2002; {\it http://
cosmologist.info/cosmomc}). This tool set allows us to work out a
suitable set of $M$ Markov Chains and analyze their statistical
properties, both in the full $\cal N$--dimensional parameter space,
and in lower dimensional subspaces. Of particular interest are the
{\it marginalized} distributions of each parameter, obtained by
integrating over the distribution of the other ${\cal N}-1$
parameters.

The above algorithms implement the following steps. (i) Let $\zeta_1$
be a point--model in the parameter space. The spectra $C_l$ of such
model are computed and their corresponding likelihood ${\cal L}_1$ is
evaluated according to
\begin{equation}
-\ln {\cal L} = \sum_l \left(l+{1 \over 2} \right) \left[{_d{C}_l|_0B_l|^2+{N}_l
\over C_l |_0B_l|^2+{N}_l}+ \ln \left({ C_l |_0B_l|^2 +{N}_l \over
_d{C}_l|_0B_l|^2 +{N}_l }\right) -1 \right] + {\rm const.} ~,
\end{equation}
where $_dC_l$ stands for either fiducial or realized spectra.
The above expression holds for a single $TT$ or $EE$ spectrum; for its
generalization to 3 spectra and a discussion of the effects of
anisotropic noise, sky cuts, etc. see, e.g. Percival \& Brown (2006).
(ii) The algorithm then randomly selects a different point $\zeta_2$,
in parameter space, according to a suitable {\it selection
  function}. The probability of accepting $\zeta_2$ is given by ${\rm
  min}({\cal L}_2 /{\cal L}_1,1)$. If $\zeta_2$ is accepted, it is
added to the chain, otherwise the multiplicity $N_1$ of $\zeta_1$
increases by 1. (iii) The whole procedure is then iterated starting
either from $\zeta_2$ or from $\zeta_1$ again, until a stopping
condition is met. The resulting chain is defined by $(N_i \zeta_i)$,
where $\zeta_i$ are the points explored and the multiplicity $N_i$ is
the number of times $\zeta_i$ was kept.

The whole cycle is repeated until the chains reach a satisfactory {\it
mixing} and a good {\it convergence}. The first requisite essentially
amounts to efficiently exploring the parameter space, by quickly
moving through its whole volume.  In particular, the fact that the
algorithm sometimes accepts points with lower likelihood than the
current point, avoids permanence in local minima, while a careful
choice of parametrization and of the selection function minimize the
time spent exploring degenerate directions.  A good convergence
instead guarantees that the statistical properties of the chains (or
of a suitably defined subset of the chains), correspond to those of
the underlying distribution.  Here we implement the convergence
criterion of Gelman \& Rubin (1992), based on $R$--ratio computation:
convergence and mixing are reached when $R-1 \ll 0.1\, $. Fulfilling
such criterion requires $N \sim 10^5$ points, in the cases considered
here.

\begin{figure}
\epsscale{.60}
\plotone{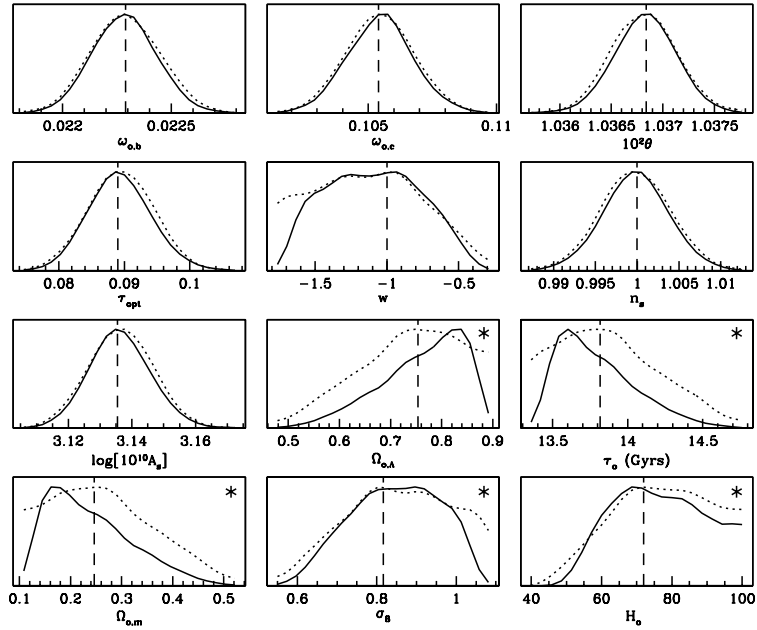}
\caption{\textsl{Marginalized {\it a posteriori} likelihood
distributions (solid lines), when artificial data derive from a
$\Lambda$CDM cosmology. Vertical lines yield the input parameter
values.  Derived parameter panels are marked by an asterisk. Dotted
lines show the average likelihood distributions. }}
\label{LCDM}
\end{figure}

The algorithm was preliminarily tested by using $\Lambda$CDM as true
cosmology and the set of parameter values shown in Figure~\ref{LCDM}.
Such Figure reports the likelihood distributions found in this case,
both for input parameters (with asterisk) and for several derived
parameters as well. Besides confirming that input values are suitably
recovered, mostly with fairly small errors, Fig.~\ref{LCDM} shows a
broad non--Gaussian distribution for $w$ values, with little skewness
but significant kurtosis. This is a well--known consequence of the
geometrical degeneracy present in CMB spectra, and clearly shows the
difficulty of CMB data to yield results on DE nature. The
non--Gaussian behavior is even more accentuated for some derived
parameter, as $\Omega_{o,c}$, $H_o$ and the Universe age $\tau_o\, .$
We shall recover analogous features in the next plots; here we want to
stress that they do not derive from fitting a partially unsuitable
parameter set.

Figure \ref{LCDM} also shows how errors increase when passing to some
secondary parameter, from primary parameters specifically devised to
break degeneracies. For instance, the errors on $\theta$ and
$\omega_{o,c}$ are $\sim 0.03\, \%$ and $\sim 1.3\, \%$, respectively.
From these parameters, $H_o$ and $\Omega_{o,c}$ are derived, whose
errors are $\sim 15\, \%$.

The point here is that the primary variable is
\begin{equation}
\label{theta}
\theta \equiv r_s(a^*)/D(a^*)~,
\end{equation}
$a^*$ setting the peak of the {\it quasi}--Gaussian last scattering
band (LSB). It is then easy to see that
$$
r_s (a^*) = {c \over H_o\, \Omega_{o,m}^{1/2}} \int^{a^*} {da \over a^2} \,
{c_s(a) \over c}\, {H_o\, \Omega_{o,m}^{1/2} \over  H(a)}~,~~
$$
\begin{equation}
\label{rs} D(a^*) = {c \over H_o\, \Omega_{o,m}^{1/2}} \int_{a^*}
^{a_o} {da \over a^2} {H_o\, \Omega_{o,m}^{1/2} \over H(a)} ~,
~~~~~~~~~~
\end{equation}
so that $\theta$ is apparently independent from $H_o$ and
$\Omega_{o,m}$ as, in the former equation
\begin{equation}
\label{hoh0} {H^2(a) \over H_o^2 ~\Omega_{o,m} } = \left(1+ {a_{eq}
\over a} \right) \left(a_o \over a \right)^3 ~,
\end{equation}
while, in the latter one
\begin{equation}
\label{hoh}
{H^2(a) \over H_o^2 ~\Omega_{o,m} } =
\left(a_o \over a \right)^3 + \left(\Omega_{o,m}^{-1}-1 \right)
g \left(a_o \over a \right)~,
\end{equation}
and $\Omega_{o,m}$ only sets the normalization of the latter term at
the r.h.s., so that $g(1) = 1\, .$ As a matter of fact, these
equations exhibit a mild dependence of $\theta$ on $\omega_{o,m}$ and
$\omega_{o,b}$, as $ a_o/a_{eq} = 2.41 \times 10^4 \omega_{o,m} (T_o/
2.726\, {\rm K})^{-4} [1.681/(1+0.227\, N_\nu)]$ and $(c_s/c)^{-2} =
3[1 + (3/4) (\omega_{o,b}/\omega_{o,m}) (a/a_{eq}) (1+0.227\, N_\nu)]$
(here $T_o$ is the present CMB temperature and $N_\nu$ is number of
neutrino families in the radiative background). Information of
$\Omega_{o,m}$ and, thence, on $H_o$ can be obtained only if the
factor $g(a_o/a)$, yielding the evolution of the ratio between DE and
matter at low redshift, is under control. It is certainly so if a
$\Lambda$CDM model is assumed; but, if we keep $w$ as free parameter,
the uncertainty on it, ranging around 60$\, \%\, ,$ reflect the
difficulty to obtain the secondary parameters $\Omega_{o,m}$
and~$H_o$.

More in general, eq.~(\ref{hoh}) shows that CMB data are sensitive to
a change of $g(a_o/a)$ causing a variation of LSB distance $D^*$.  On
the contrary, $\theta$ does not discriminate between different $w(a)$
yielding the same $D^*$.

Incidentally, all that confirms that it is unnecessary to fix $a^*$
with high precision, or to discuss whether it is a suitable indicator
of the LSB depth. Knowing $a^*$ with $\sim 1\, \%$ approximation is
quite suitable to this analysis.

\section{Dynamical DE vs.~$w={\rm const.}$}
Let us now discuss what happens if DE state equation, in the ``real''
cosmology, cannot be safely approximated by a constant $w\, .$ To
build data, we use then a SUGRA cosmology and our own extension of
CAMB, directly dealing with a SUGRA potential both in the absence and
in the presence of DM--DE coupling. Artificial data worked out for
Model A with $\beta=0$ were then fit to the same parameters as in
Figure \ref{LCDM}. The whole findings are described in Table
\ref{tab1} (second column).

\begin{figure}
\epsscale{.60}
\plotone{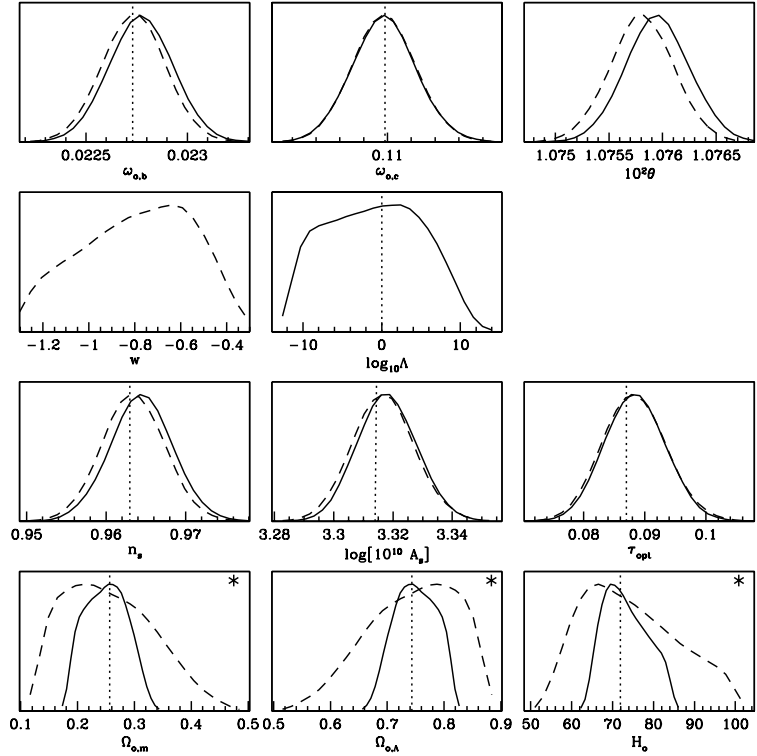}
\caption{\it{Marginalized likelihood distributions on parameter
values. Data built with an uncoupled SUGRA model. The parameter space
for the fit includes either $\log(\Lambda/{\rm GeV})$ (solid lines) or
a constant $w$ (dashed lines). Vertical dotted lines show the input
parameter values.  Derived parameters plots are with asterisk.  }}
\label{dde}
\end{figure}

In Figure \ref{dde} we add further information, comparing the value
distributions in fits assuming either a cosmology with $w={\rm
const.}$ or a SUGRA cosmology, then including $\Lambda,$ as parameter,
instead of $w$. The two fits agree well within 1--$\sigma$, among
themselves and with input values. In the Figure we show the fiducial
case.

\begin{table}[!hbt]
\vskip -.2truecm
\begin{center}
\begin{tabular}{cccc}
\hline
\multicolumn{4}{c}{{~~~~~~~~~~~~~~~~~~~~~
\rm Input~model:~{\it SUGRA} ($\Lambda = 1\, $GeV)~with}} \\
{\rm Parameter} \&  & $\beta=0$ & $\beta=0.05$ & $\beta=0.1$
\\
 {\rm Input~value} & 
{\rm Av.~value}~$\pm\sigma$ & {\rm Av.~value}~$\pm\sigma$ & {\rm Av.~value}~$\pm\sigma$ 
\\
\hline
$10^2 \, \omega_{o,b}$  &$2.274\pm0.015$ &$2.274\pm0.015$ &$2.277\pm0.017$\\
$2.273$                 &$2.278\pm0.015$ &$2.275\pm0.015$ &$2.295\pm0.017$\\
                        &$2.278\pm0.015$ &$2.261\pm0.015$ &$2.287\pm0.017$\\
                        &$2.280\pm0.015$ &$2.278\pm0.015$ &$2.282\pm0.017$\\
\hline
$\omega_{o,c}$      &$0.1099\pm0.0013$ &$0.1164\pm0.0014$ &$0.1225\pm0.0016$\\
 $0.1099$           &$0.1083\pm0.0013$ &$0.1166\pm0.0012$ &$0.1225\pm0.0016$\\
                    &$0.1086\pm0.0013$ &$0.1171\pm0.0014$ &$0.1216\pm0.0016$\\
                    &$0.1104\pm0.0013$ &$0.1163\pm0.0014$ &$0.1225\pm0.0015$\\
\hline
$10^2 \, \theta$    &$1.0758\pm0.0003$ &$1.0736\pm0.0003$ &$1.0507\pm0.0003$\\
 $1.072$            &$1.0759\pm0.0003$ &$1.0736\pm0.0003$ &$1.0509\pm0.0003$\\
                    &$1.0760\pm0.0003$ &$1.0737\pm0.0003$ &$1.0507\pm0.0003$\\
                    &$1.0759\pm0.0003$ &$1.0736\pm0.0003$ &$1.0507\pm0.0003$\\
\hline
$\tau_{opt}$        &$0.088\pm0.005$ &$0.087\pm0.005$ &$0.085\pm0.005$ \\
 $0.087$            &$0.087\pm0.005$ &$0.903\pm0.005$ &$0.084\pm0.005$\\
                    &$0.089\pm0.005$ &$0.083\pm0.005$ &$0.087\pm0.005$\\
                    &$0.093\pm0.005$ &$0.079\pm0.005$ &$0.078\pm0.005$\\
\hline
$w$                 &$-0.79-0.12+0.49$ &$-0.75-0.30+0.26$ &$-0.85-0.43+0.38$\\
---                 &$-0.84-0.27+0.25$ &$-0.76-0.30+0.26$ &$-0.96-0.44+0.82$\\
                    &$-0.81-0.26+0.25$ &$-0.79-0.29+0.26$ &$-0.67-0.34+0.37$\\
                    &$-0.87-0.28+0.28$ &$-0.54-0.15+0.16$ &$-0.63-0.24+0.23$\\
\hline
$n_s$               &$0.963\pm0.004$ &$0.962\pm0.004$ &$0.960\pm0.004$ \\
 $0.963$            &$0.966\pm0.004$ &$0.962\pm0.004$ &$0.958\pm0.004$\\
                    &$0.968\pm0.004$ &$0.962\pm0.004$ &$0.963\pm0.004$\\
                    &$0.959\pm0.004$ &$0.961\pm0.004$ &$0.959\pm0.004$\\
\hline
$\ln(10^{10}\, A_s)$ &$3.3168\pm0.0102$&$3.1695\pm0.0101$ &$2.8860\pm0.0101$ \\
 $3.3144$~for~$\beta=0$ \hfill &
$3.3127\pm0.0094$ &$3.1699\pm0.0106$ &$2.8866\pm0.0099$ \\
 $3.1634$~for~$\beta=0.05$ \hfill &
$3.3144\pm0.0103$&$3.1565\pm0.0095$&$2.8885\pm0.0102$ \\
 $2.8902$~for~$\beta=0.10$ \hfill&
$3.3310\pm0.0097$&$3.1486\pm0.0105$ &$2.8736\pm0.0111$ \\
\hline
$100 \, h$          &$74.7-7.2+25.3$  &$70.6-11.3+12.8$ &$66.3-11.7+13.8$ \\
 $71.9$             &$77.3-12.4+13.7$ &$71.0-11.4+12.6$ &$70.2-20.3+14.6$\\
                    &$75.7-12.3+13.3$ &$71.9-11.0+12.6$ &$60.9-11.1+10.5$\\
                    &$78.4-13.6+13.9$ &$61.6-6.1+6.0$   &$59.5-5.7+5.6$\\
\hline
$\Omega_{o,m}$ &$0.255{-0.084}{+0.169}$ &$0.300{-0.100}{+0.095}$ 
&$0.360{-0.133}{+0.127}$\\
 $0.257$       &$0.235{-0.105}{+0.077}$ &$0.296{-0.096}{+0.096}$ 
&$0.325{-0.123}{+0.140}$\\
               &$0.245{-0.079}{+0.077}$ &$0.289{-0.093}{+0.088}$ 
&$0.422{-0.138}{+0.121}$\\
               &$0.233{-0.101}{+0.084}$ &$0.379{-0.074}{+0.073}$ 
&$0.431{-0.103}{+0.097}$\\
\hline
\end{tabular}
\vskip -.1truecm
\caption{\it Results of an MCMC analysis, seeking the parameters
listed in the first column, on artificial CMB data built with the
parameter values also listed in the first column, but using a SUGRA
cosmology, whose $\Lambda$ and $\beta$ are shown in the header. For
each parameter, the first line yields results for the fiducial case,
the next 3 lines for model realizations.}
\label{tab1}
\end{center}
\end{table}

Let us briefly comment the results in the absence of coupling, when
the fitting parameter set includes $w$ instead of $\Lambda$: (i)
$\omega_{o,b}$ and $\omega_{o,c}$ are recovered without any bias. (ii)
The same holds for $\theta,$ if a numerically refined value of $a^*$
is used; the expression of $a^*$, as above outlined, has a precision
$\sim 1\, \%$. Errors on $a^*$, however, are negligible in comparison
with errors arising from the scarce knowledge of $w(a)$.

As outlined in the previous Section, the main problem to fix
$\Omega_{o,m}$, $\Omega_{o,\Lambda}$, and $H_o$, resides in the
difficulty to determine the state equation of the expansion source
since DE becomes substantially sub--dominant or dominant. This also
makes clear why the formal error in the determination of the above
three parameters is wider when the fit assumes a constant $w:$ varying
$w$ yields an immediate and strong effect on DE contributions and
state equation, extending up to now; on the contrary, only huge
displacements of the energy scale $\Lambda$ cause significant
variations to DE contributions and state equation within the family of
SUGRA cosmologies. In other terms, when $\Lambda$ varies, the part of
the functional space that $w(a)$ covers is not so wide.

Accordingly, the greater error bars on $\Omega_{o,m}$,
$\Omega_{o,\Lambda}$, $H_o$, found when we use $w$ as a parameter, do
not arise because we fit data to a ``wrong'' cosmology, but because
varying $w$ leads to a more effective spanning of the functional space
of $w(a)$.

As a matter of fact, the reliability of the fit must be measured by
comparing the likelihood of the best--fit parameter sets. Likelihood
values, given in the next Section, do not discriminate between the two
fits. Otherwise, a direct insight into such reliability is obtained by
looking at the width of the marginalized {\it a posteriori} likelihood
distributions on primary parameter values. Figure \ref{dde} then
confirms that such distributions, although slightly displaced, have
similar width for all primary parameters.

A general conclusion this discussion allows to draw is to beware from
ever assuming that error estimates on secondary parameters, as
$\Omega_{o,m}$, $\Omega_{o,b}$, and $H_o$, are safe.

\begin{figure}
\epsscale{.55}
\plotone{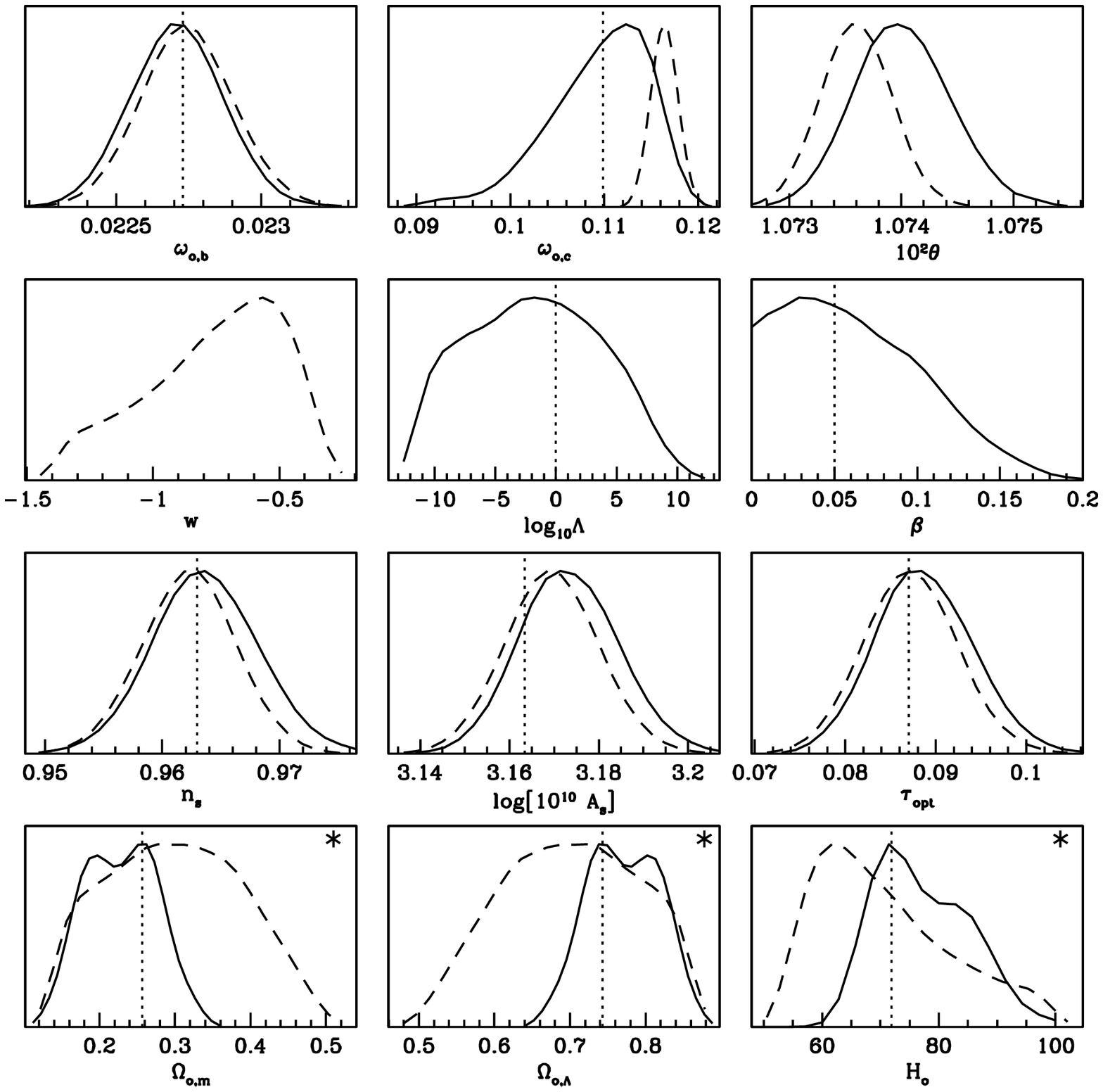}
\vskip1.2truecm
\plotone{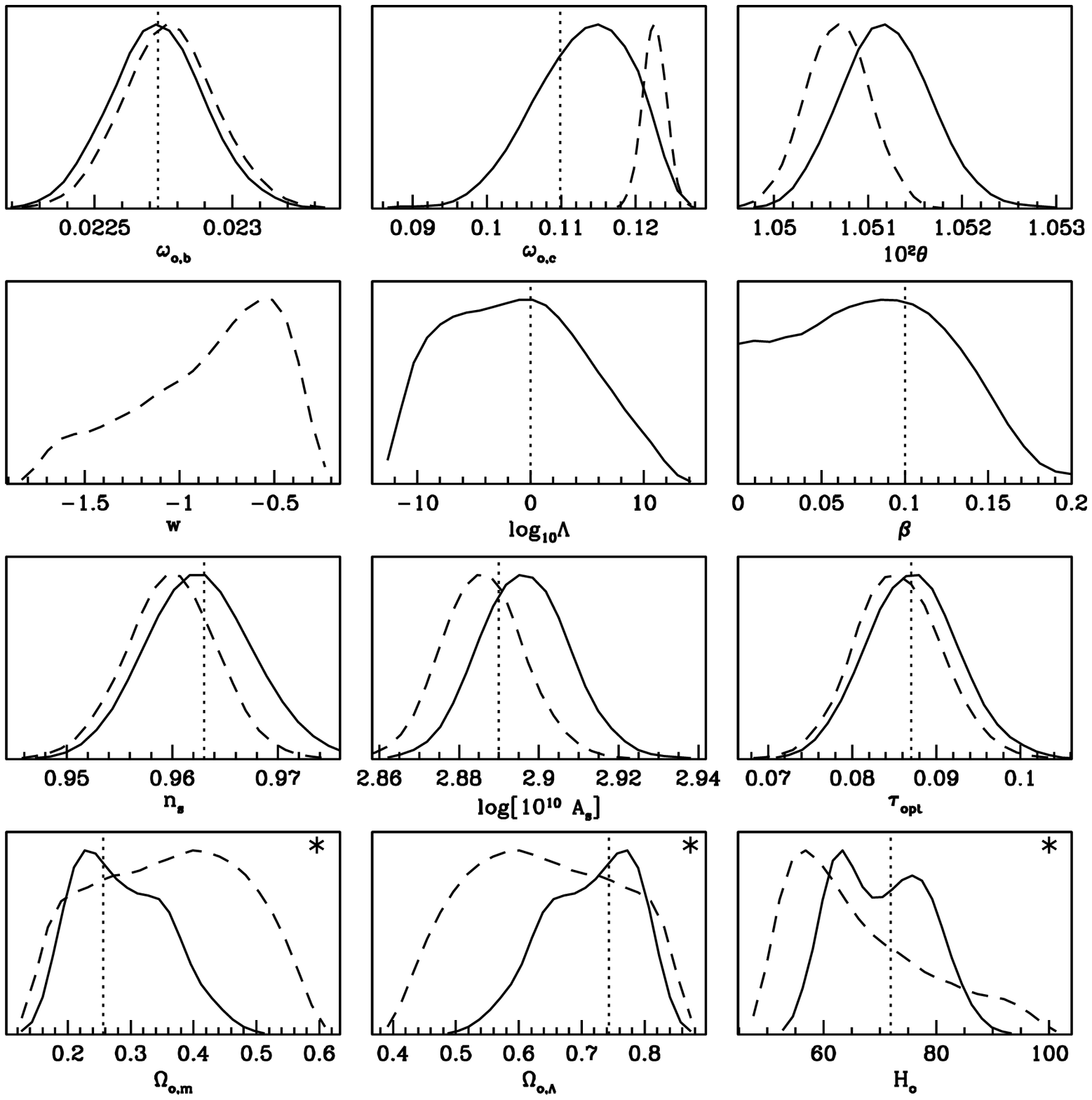}
\caption{\it{Marginalized likelihood distributions on parameter
values. Data built with coupled SUGRA models with $\beta = 0.05$,
$\beta = 0.1$ (upper, lower panels). The parameter space for fits
includes $\log(\Lambda/{\rm GeV})$ and $\beta$ (solid lines) or just a
constant $w$ (dashed lines). Vertical dotted lines show the input
parameter values.  Pay attention to the different abscissa units in
the two panel sets. }}
\label{couple}
\end{figure}

\section{Coupled DE vs.~$w={\rm const.}$}

In cosmologies including cold DM and DE, the equation obeyed by
the stress--energy tensors of the dark components
\begin{equation}
\label{conti0}
T^{(c)~\mu}_{~~~~\nu;\mu} + T^{(de)~\mu}_{~~~\, ~~\nu;\mu} = 0~,
\end{equation}
yields
\begin{equation}
{d \over d\tau} (\rho_c + \rho_{de}) = -3 (\rho_c + p_c +
\rho_{de} + p_{de}) {\dot a \over a} ~, \label{conti}
\end{equation}
if their pressure and energy densities are $p_{c}$, $p_{de}$ and
$\rho_{c},$ $\rho_{de};$ $\tau$ is the conformal time and dots
indicate differentiation in respect to $\tau.$ Eqs.~(\ref{conti0}) and
(\ref{conti}) state that no force, apart gravity, acts between
standard model particles and the dark components. The
eq.~(\ref{conti0}) is fulfilled if DM and DE separately satisfy the
eqs.
\begin{equation}
T^{(de)~\mu}_{~~~\, ~~\nu;\mu} = C T^{(c)} \phi_{,\nu}~,~~~~~~~~
T^{(c)~\mu}_{~~~~\nu;\mu} =- C T^{(c)} \phi_{,\nu}~,
\end{equation}
where $T^{(de),(c)}$ are traces of the stress--energy tensors. $C$ can
be an arbitrary constant or, {\it e.g.}, a function of $\phi$ itself;
when $C=0$, the two dark components are uncoupled and so are the cases
we considered up to here.

It is however known (Wetterich 1995, Amendola 2000, Amendola \&
Quercellini 2003) that self consistent theories can be built with
\begin{equation}
C \equiv 4 \sqrt{\pi \over 3} { \beta \over m_p} \neq 0~,
\end{equation}
so that $\beta $ is used to parametrize the strength of DM--DE
coupling. 

By using MCMC techniques, we explored cosmologies with $\beta = 0.05$
and 0.1 (Model A); we also report some results for $\beta = 0.02$ (Model
B). We aim to show that, when $\beta$ is excluded from the parameter
budget, the values MCMC provide for some parameters can be
significantly biased, even though $\beta $ is small.

\begin{figure}
\epsscale{.60}
\plotone{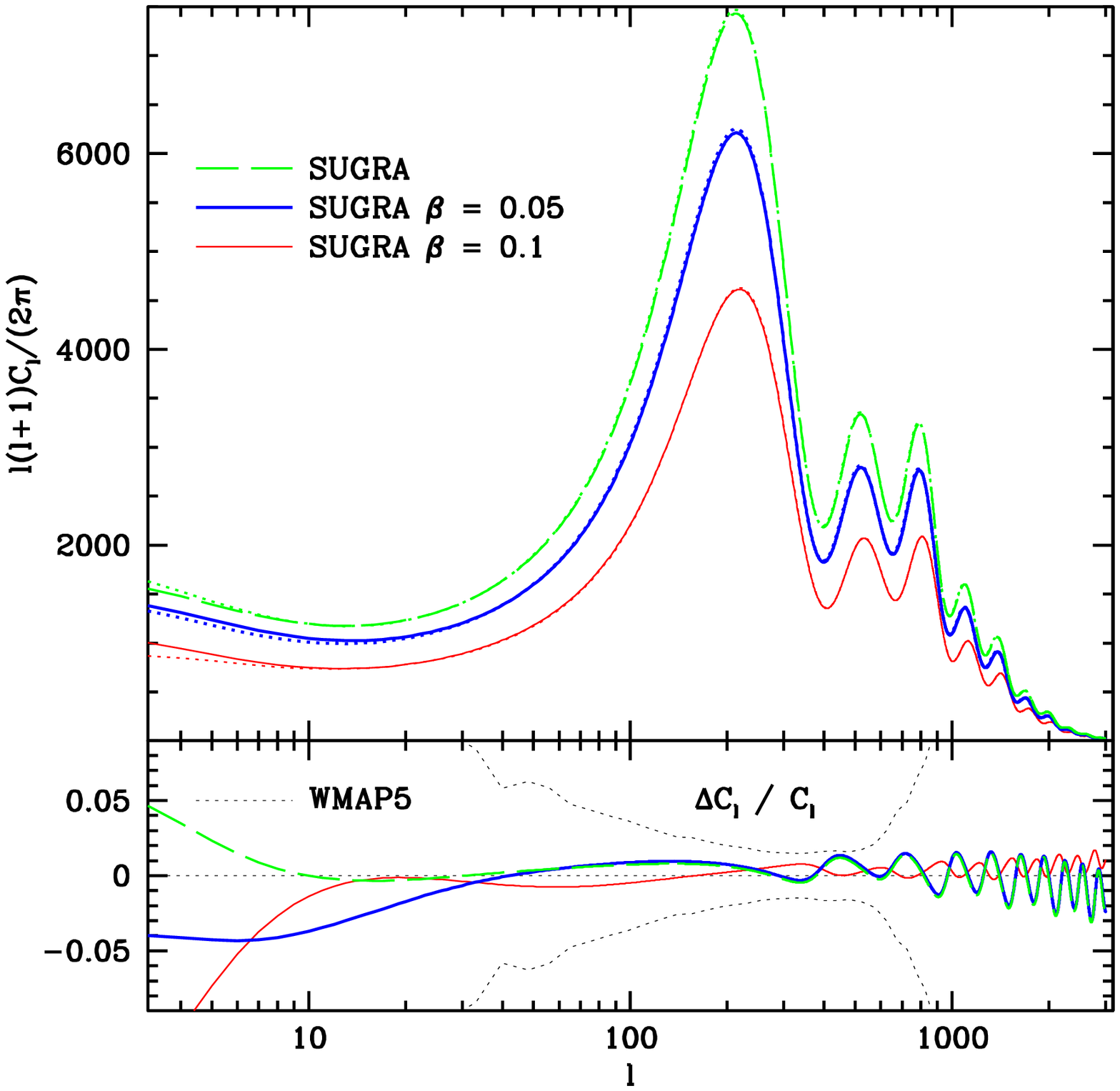}
\caption{\it Anisotropy spectra comparison. The 3 values of $\beta$
considered are clearly distinguished thanks to their different input
normalization and to the line type indicated inside the upper frame.
The dotted lines (hardly visible at low $l$'s in the upper panel) are
the spectra of the corresponding best--fit model when assuming no
coupling and $w = const.\, $. In the lower panel the relative
differences between input and best--fit models are compared with WMAP5
error~size.  }
\label{Cla}
\end{figure}
\begin{figure}
\epsscale{.60}
\plotone{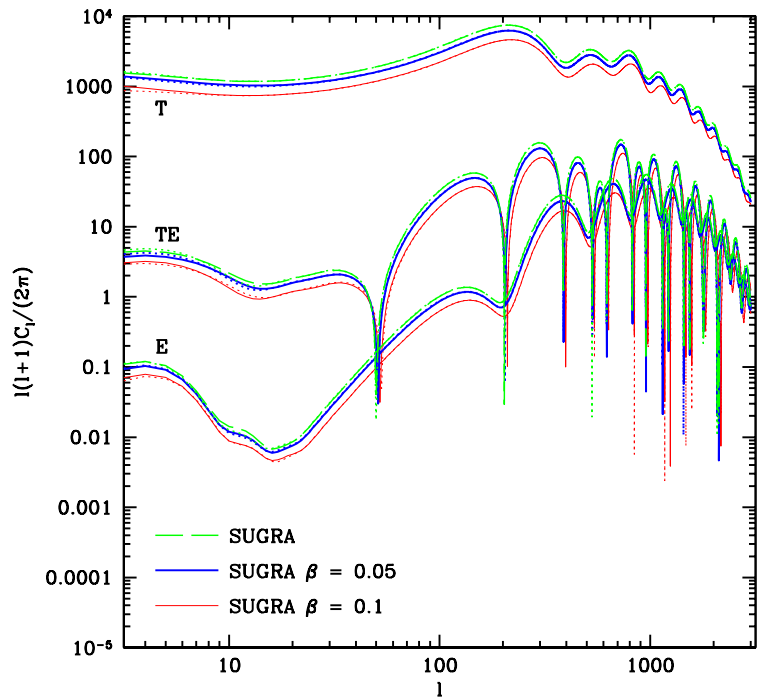}
\caption{\it The comparison made in the upper panel of the previous
Figure is extended to ET and E--polarization spectra. Model differences
are hardly visible at very low $l$ and on the $l$ values where the
ET spectrum changes sign. }
\label{Clb}
\end{figure}

\begin{figure}
\epsscale{.50}
\plotone{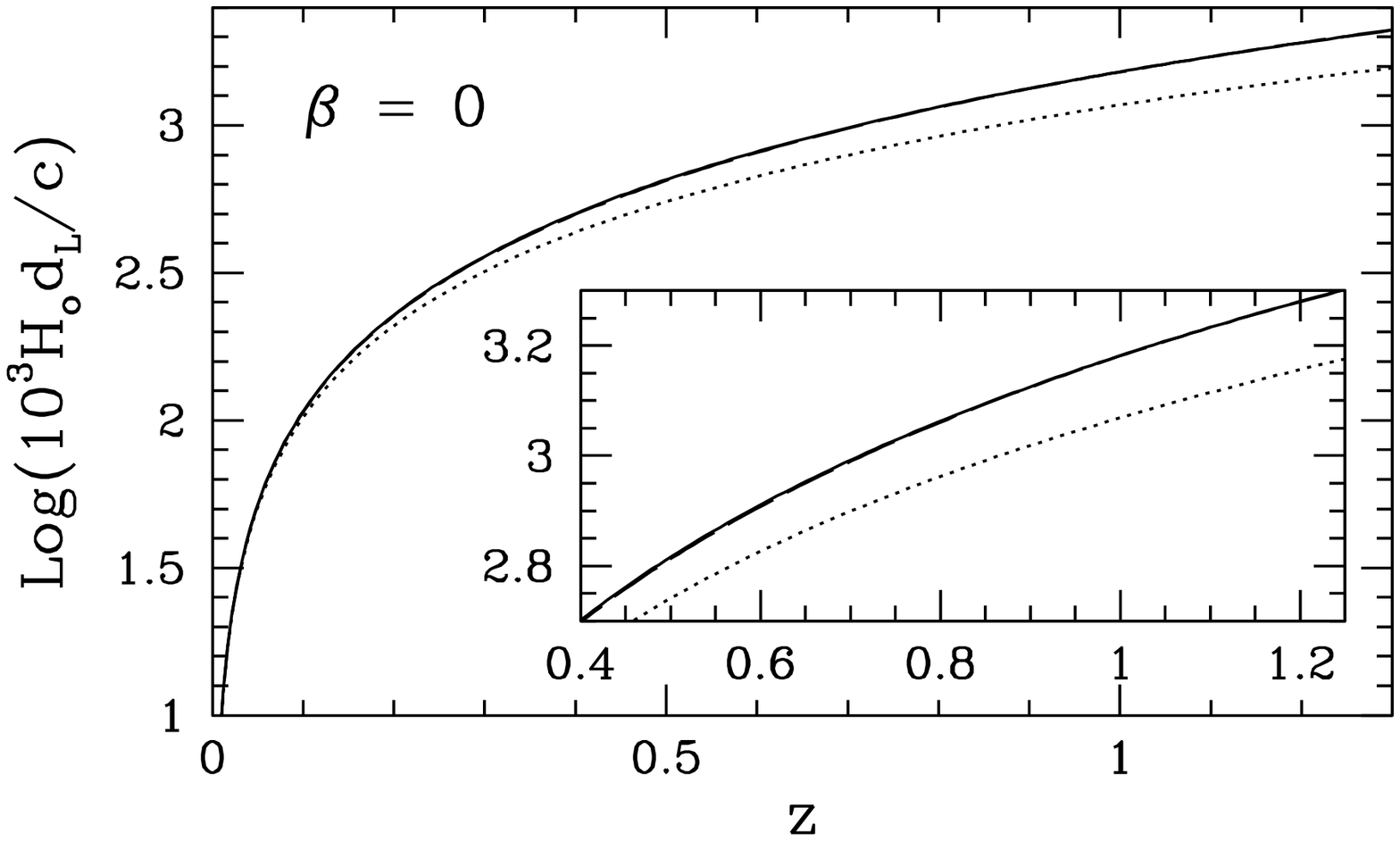}
\vskip0.5truecm
\plotone{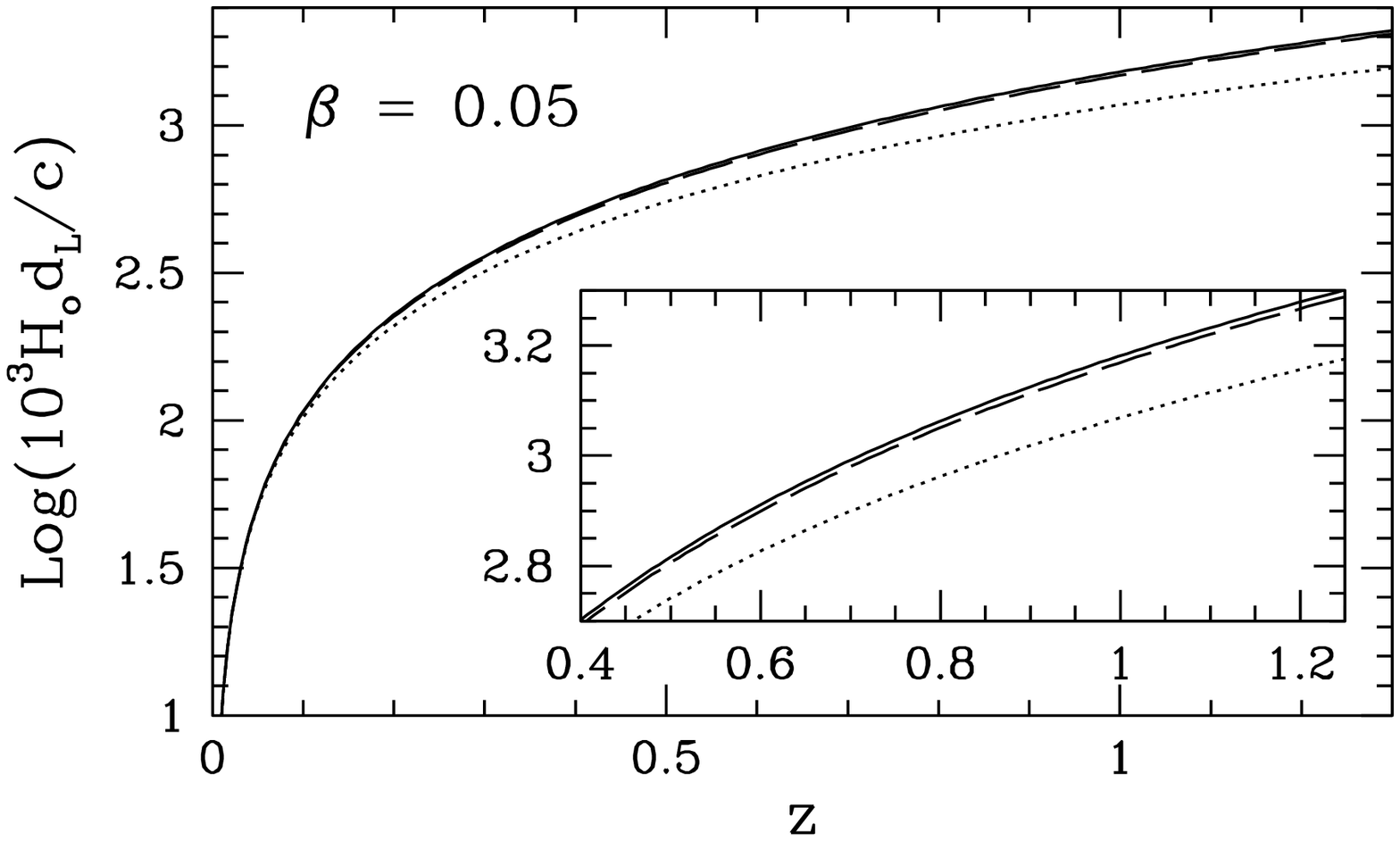}
\vskip0.5truecm
\plotone{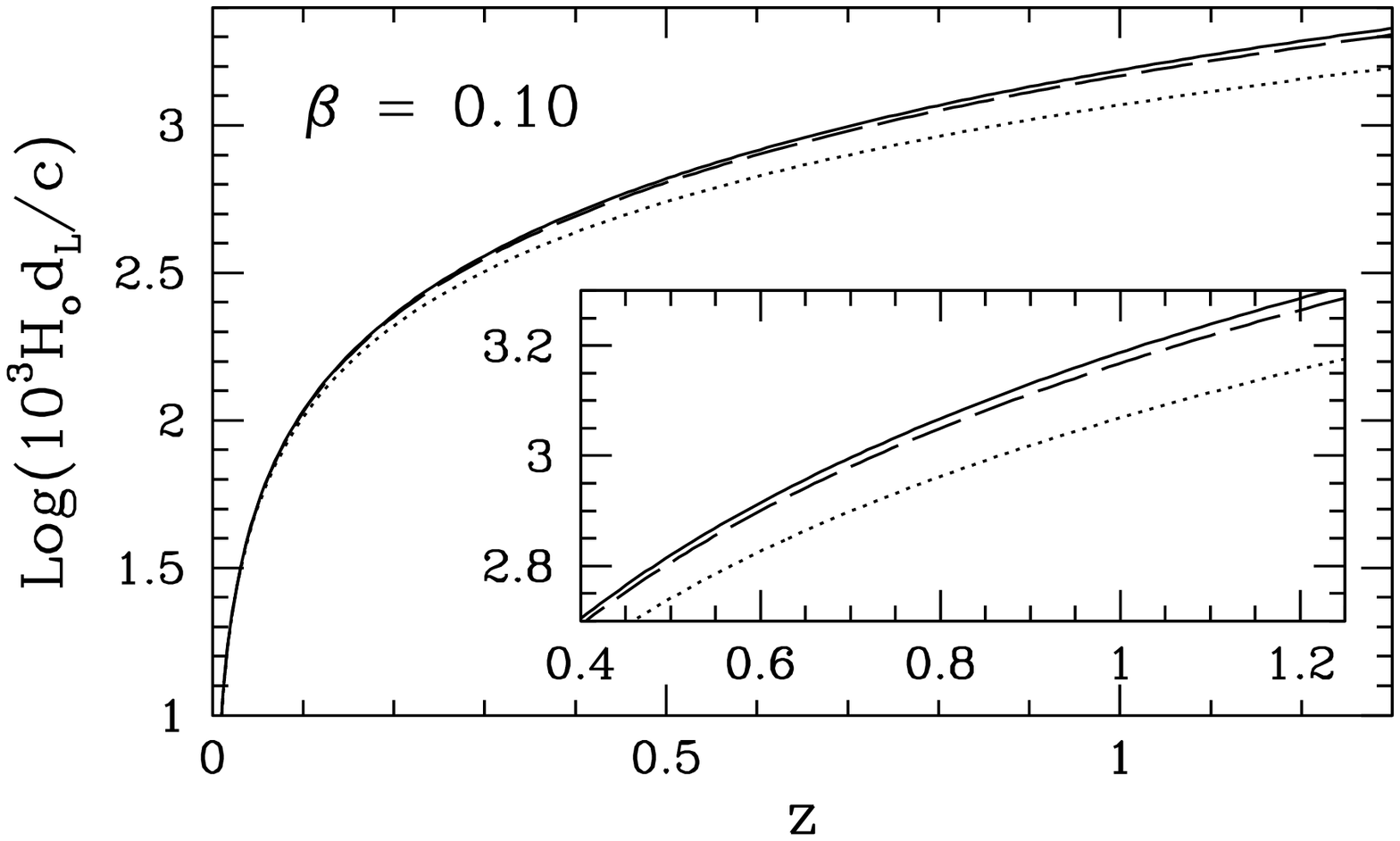}
\caption{\it Luminosity distance vs.~redshift in models with $\beta =
$0, 0.05, 0.10~(solid lines).  Each model is compared with the
corresponding best--fit constant--$w$ model (dashed line) and
with a SCDM model (dotted line) with equal $\Omega_b~.$
}
\label{hubdia}
\end{figure}
\begin{figure}
\plotone{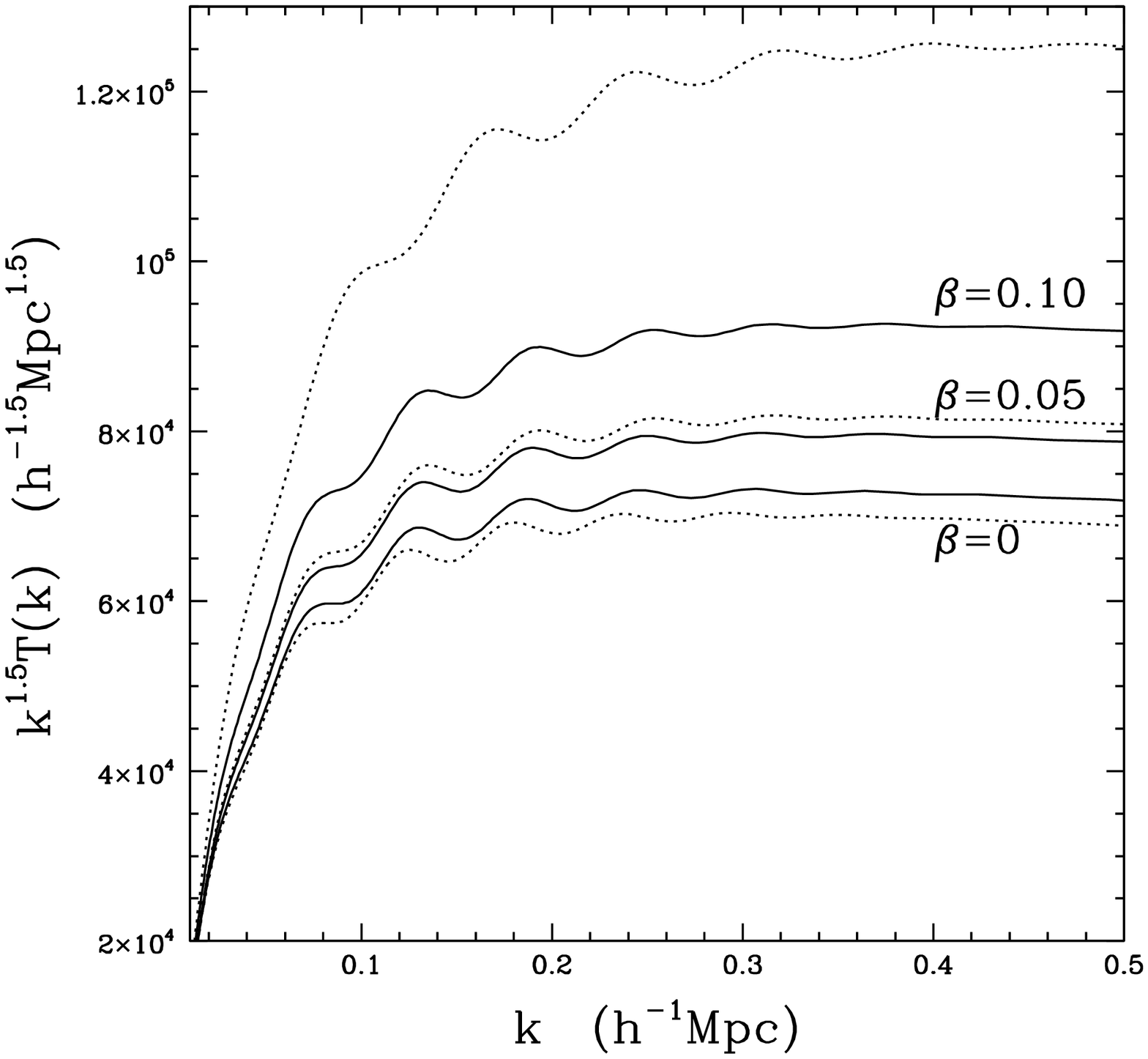}
\caption{\it Transfer functions for the models A. The dotted lines
yield the transfer function of the models with w=const., yielding
the best--fit for CMB spectra. }
\label{trans}
\end{figure}

In Table \ref{tab1} (3rd--4th columns) we report the results of
fitting the parameter set including constant $w$, in place of
$\Lambda$ and $\beta$. The Table allows an easy comparison with the
uncoupled case, when all parameters unrelated to DE are nicely
recovered. On the contrary, already for $\beta = 0.05$, the input
$\omega_{o,c}$ value is formally $4.5~\sigma$'s away from best--fit,
in the fiducial case, and more than 5~$\sigma$'s in some realization.
The effect is stronger for $\beta = 0.1$, yielding a discrepancy $\sim
8$ $\sigma$'s. That coupling affects the detection of $\omega_{o,c}$
is not casual: in coupled models a continuous exchange of energy
between DM and DE occurs.

Several parameters exhibit a non--Gaussian distribution, however.
This is visible in Figure \ref{couple}, where we also provide a
comparison between likelihood distributions when (i) the fit includes
just $w$ or, instead, (ii) the parameter space includes $\Lambda$ and
$\beta$.

The comparison allows to appreciate that, in the (i) case,
$\omega_{o,c}$ is apparently much better determined than in the (ii)
case. Appreciable discrepancies concern also the $\theta$ parameter and
they reflect onto the derived parameters $\Omega_{o,b}$,
$\Omega_{o,m}$ and $H_o$, whose distribution however appears
significantly non--Gaussian.

These results deserve to be accompanied by a comparison between the
likelihood values in the different cosmologies. The following table
shows 
$\chi^2_{\rm eff} \equiv -2 \, {\rm ln}
({\cal L}) $ for the fiducial cases only:
$$
\left|
\matrix{
& & {\rm constant~} w & \Lambda~\&~\beta \cr
{\bf }~~ & & ----- & ----- \cr
{\bf }~~ &\beta = 0 \hfill & 0.175  & 0.887 \cr
{\bf }~~ &\beta = 0.05 \hfill & 0.214 & 0.332 \cr
{\bf }~~ &\beta = 0.10 \hfill & 0.778 & 0.251 }
\right|
$$
The likelihood values for realizations are systematically smaller, as
expected, but confirm the lack of significance shown here. Given that
coupled models have 1 additional parameter over the $ w = const$
model, differences $\Delta \chi^2_{\rm eff} < \sim 1$ indicate that
both models are an equally good fit of data. The 
conclusion is that spectral differences, between input and best--fit
models, 
lay systematically below the error size. As a matter of fact, we are
apparently meeting a case of degeneracy.

This statistical observation is corroborated by a direct insight into
angular spectra, provided by Figures \ref{Cla} and \ref{Clb}.  We show
the behaviors of the $C_l$ spectra for the Model A, with $\beta=0$,
$\beta=0.05$, $\beta=0.1$, compared with the $C_l$ for the
best--fitting ($w={\rm const.}$) model. Differences are so small to be
hardly visible. Figure \ref{Cla} refers just to anisotropy. In its
lower panel the $\Delta C_l/C_l$ ratio is compared with WMAP5
1--$\sigma$ error size ($\Delta C_l$ is the difference between the
spectra of the input and best--fit models). Notice that shifts
by 1 or 2 units along the abscissa,
at large $l$ values,
would still badly cut--off the apparent
ratios. Clearly, degeneracy can be removed only if the error size is
reduced by a factor $\sim 10$ or more, in the region between the first
deep and the second acoustic peak.

A further example we wish to add concerns a still smaller coupling
intensity, $\beta = 0.02$ (Model B, Table \ref{tab02}). At this
coupling level, the MCMC meet all input parameter values within a
couple of $\sigma$'s. A closer inspection of plots similar to Figure
\ref{couple}, not reported here, shows that the probability of
realizations yielding $\omega_{o,c}$ formally more than 3--$\sigma$'s
away from the true value, is still $> 8 \, \%:$ the genie is still not
completely back inside the lamp.

\begin{table}
\begin{center}
\begin{tabular}{cc}
\hline \hline
\multicolumn{2}{c}{{\rm Input~model:~coupled {\it SUGRA} ($\Lambda = 1\, $GeV,~~$\beta=0.02$)}} \\
\hline \hline {\rm Parameter} \& {\rm Input~value} & {\rm
b.f.~value}~$\pm\sigma$
\\
\hline \hline
$10^2 \, \omega_{o,b} = 2.400$               & $2.400\pm0.016$ \\
                                                   & $2.424\pm0.016$ \\
                                                   & $2.423\pm0.017$ \\
                                                   & $2.391\pm0.016$ \\
\hline
$\omega_{o,c} = 0.1100$              & $0.1127\pm0.0014$ \\
                                                   & $0.1120\pm0.0013$ \\
                                                   & $0.1121\pm0.0013$ \\
                                                   & $0.1130\pm0.0014$ \\
\hline
$10^2 \, \theta$                               & $1.11135\pm0.00032$ \\
                                                   & $1.11174\pm0.00031$ \\
                                                   & $1.11170\pm0.00032$ \\
                                                   & $1.11108\pm0.00032$ \\
\hline
$\tau = 0.1500$              & $0.1515\pm0.0059$ \\
                                                   & $0.1413\pm0.0067$ \\
                                                   & $0.1412\pm0.0061$ \\
                                                   & $0.1510\pm0.0061$ \\
\hline
$w$ (at $z=0$)                         & $-0.70-0.16+0.16$ \\
                     &            $-0.65-0.17+0.17$ \\
                                                   & $-0.67-0.17+0.16$ \\
                                                   & $-0.63-0.14+0.13$ \\
\hline
$n_s = 1.0000$              & $0.9998\pm0.0042$ \\
                                                   & $0.9994\pm0.0039$ \\
                                                   & $0.9990\pm0.0041$ \\
                                                   & $0.9976\pm0.0041$ \\
\hline
$\ln{(10^{10} \, A_s)} = 3.136$               & $3.138\pm0.011$ \\
                                                   & $3.119\pm0.013$ \\
                                                   & $3.119\pm0.012$ \\
                                                   & $3.139\pm0.012$ \\
\hline
$100 \, h = 85$                      & $79-9\phantom{0}+10$ \\
                                                   & $77-10+10$ \\
                                                   & $77-10+10$ \\
                                                   & $75-8\phantom{0}+8\phantom{0}$ \\
\hline
$\Omega_{o,m} = 0.185$               & $0.254-0.057+0.058$ \\
                                                   & $0.294-0.062+0.062$ \\
                                                   & $0.268-0.059+0.059$ \\
                                                   & $0.219-0.054+0.054$ \\
\hline\hline
\end{tabular} 
\caption{\textsl{As previous Table, with different input values of
cosmological parameters and $\beta$ as small as 0.02}}
\label{tab02}
\end{center}
\end{table}

\section{Other observables}
The degeneracy observed in CMB spectra could be broken off through
different observables. We plan to deepen this aspect in a forthcoming
work, by building detailed data sets accounting from the dependence on
cosmology of the expansion rate and growth factor.

A first insight into the actual situations can be however gained
through an inspection of Hubble diagrams and transfer functions.

In Figure \ref{hubdia} we show the redshift dependence of the
luminosity distances for models with $\beta=0$, 0.05, 0.1 and compare
them with the corresponding best--fit models with constant $w$, as
well as with SCDM models with the same value of $\Omega_b$.  These
plots clearly indicate that a fit with SNIa data would hardly allow
any discriminatory signal: discrepancies from constant--$w$ model
increase with $\beta$; but, even for the $\beta = 0.10$ case, they 
hardly exceed $\sim 10\, \%$ of the difference from SCDM.

In Figure \ref{trans} we then exhibit the transfer functions $T(k)$
(multiplied by $k^{1.5}$ to improve the visibility of details) for
models A.

In the cases $\beta = 0$ and $\beta = 0.05$, we notice slight
displacements for the BAO system and the slope. The actual setting of
BAO's is however subject to non--linear effects and residual
theoretical uncertainties are wider than the shifts in the plots.  The
change of slope is also easily compensated by a shift of $n_s$ by
$\sim 0.01$, widely within expected observational errors.  Notice then
that the relative position of the input and best--fit functions is
opposite in the two cases. Accordingly, in the intermediate case
$\beta = 0.033$, not shown in the plot, the overlap is almost exact
and the observed scale dependence of the growth factor, at low
redshift, does not break the degeneracy at all.

The situation is different when $\beta > 0.05$ is considered. For
$\beta = 0.1$, the BAO displacement is indeed relevant; to compensate
for the change of slope, we would then require a shift of $n_s$
greater than 0.1$\, $.

A numerical analysis can therefore determine at which value, probably
intermediate between 0.05 and 0.1, the CMB spectra degeneracy is
broken.

\section{Conclusions}
The nature of the dark cosmic components is unclear. DE could be a
self--interacting field, yielding a scale dependent state parameter
$w(a)\, .$ Which bias does then arise on cosmological parameter
estimates, if performed by assuming $w = {\rm const.}\, $?  This
question regards also parameters which do not describe DE; their
estimate could be biased because the {\it true} model is not directly
explored. A first conclusion of this work is that such bias exists
but, in the cases we treated, yields acceptable displacements, within
1--$\sigma$.

Suppose however that future data allow to exclude $n_s = 1$, within
3--$\sigma$'s, when we assume $w = {\rm const.}\, $. Setting $n_s < 1$
discriminates among inflationary potentials. It would be however
legitimate to assess that, at that stage, such conclusions would still
be premature.

Our analysis was then extended to the case of DM--DE coupling. The
idea that DM and DE have related origins or arise from the same field
has been widely pursued (see, {\it e.g.}  Kamenshchik, Moschella \&
Pasquier 2001, Bento, Bertolami \@ Sen 2002, 2004; Mainini \&
Bonometto 2004; for a review, see~Cop\-eland, Sami \& Tsujikawa
2006). Coupling causes DM--DE energy exchanges and this option was
first explored in the attempt to ease the {\it coincidence} problem.

Unfortunately, when a DM--DE coupling, strong enough to this aim, is
added to models, predictions disagree with data. This does not forbid,
however, that the physics of the dark components includes a weaker
coupling or that a stronger coupling is compensated by other features
(see, {\it e.g.}, La Vacca et al. 2008).

In this work we pointed out that a significant degeneracy exists, so
that we can find an excellent fit of CMB data for coupled DE
cosmologies just by using constant--$w$ uncoupled cosmologies. The fit
is so good that even likelihood estimates do not allow to distinguish
between the ``true'' cosmology and the best--fit constant--$w$ model,
at the present or foreseeable sensitivity levels. {\it Unfortunately,
however, the values obtained for several parameters are then widely
different from input ones.}

In fact, if we ignore the coupling degree of freedom, when data are
analyzed, we can find biased values for some primary parameter as
$\omega_{o,c}$, for which input values lay $\sim 5$--$\sigma$'s away
from what is ``detected''.  Also $\Omega_{o,m} $ and $H_o$, which are
secondary parameters, are significantly biased. In particular, with a
coupling as low as $\beta = 0.05$, we found model realisations
yielding $H_o$ estimate $\sim 2$--$\sigma$'s away from the input
value. If we keep to the fiducial case, however, the probability to
find $H_o \geq $ its input value is 36.5$\, \%$ for $\beta = 0.05$ and
26.5$\, \%$ for $\beta = 0.1~.$ This outlines the tendency to find
smaller values than ``true'' ones.

As a matter of fact, however, because of the uncertainty induced by
our ignorance on DE state equation, the width of errors on secondary
parameter significantly exceeds the width for primary ones, and a
5--$\sigma$ discrepancy is partially hidden by such ignorance. This
agrees with the fact that data analysis shows a significant increase
of errors on secondary parameters when the set of models inspected
passes from $\Lambda$CDM to generic--$w$ cosmologies.  For instance,
in the very WMAP5 analysis, the error on $H_0$ increases by a huge
factor $\sim 6$, when one abandons the safety of $\Lambda$CDM to
explore generic constant--$w$ models. A general warning is then that
errors obtained assuming $\Lambda$CDM are to be taken with some
reserve.

An important question is whether the bias persists when different
observables, besides CMB, are used. A preliminary inspection shows
that SNIa data would hardly provide any discrimination. On the
contrary an analysis of fluctuation growth can be discriminatory if
$\beta >\sim 0.07$--0.08~.

An even more discriminatory signal could be found in the $z$
dependence of the growth factor is suitably analyzed.  Observational
projects aiming at performing a tomography of weak lensing, like
DUNE--EUCLID (see, {\it e.g.}, Refre\-gier et al. 2006, 2008),
therefore, can be expected to reduce this degeneracy~case.

\vskip .3truecm

\noindent
ACKNOWLEDGMENTS. LPLC is supported by NASA grant NNX07AH59G and JPL--Planck
subcontract no. 1290790.

\vfill\eject


\begin{thebibliography}{20}

\bibitem{A1} Amendola L. (2000) PR D{\bf 62}, 043511.

\bibitem{A2} Amendola L. \& Quercellini C. (2003)  PR D{\bf 68},
023514.

\bibitem{A4} Amendola L., Campos G.C. \& Rosenfeld R. (2004)
astro-ph/0410543.

\bibitem{A3} Amendola L., Quercellini C., Tocchini-Valentini D. \&
Pasqui A. (2003)  ApJ Lett.~{\bf 583}, 53.

\bibitem{B1} Brax P. \& Martin J. (1999) PL B{\bf 468},  40;
(2001) PR D{\bf 62}, 10350.

\bibitem{B2} Brax P., Martin J. \& Riazuelo A. (2000) PR D{\bf
61}, 103505.

\bibitem{B0} Bento M.C., Bertolami O. \& Sen A.A. (2002) PR D{\bf 66}, 043506;
(2004)  PR D{\bf 70}, 107304.

\bibitem{C0} Copeland E.J., Sami M. \& Tsujikawa S. (2006)
Int.J.Mod.Phys. D{\bf 15}, 1753.

\bibitem{C1} Colombo L.P.L. \& Gervasi M. (2007) JCAP {\bf 0610}, 001.

\bibitem{F1} Foley R.J. et al. (2007) astro-ph/0710.2338.

\bibitem{??} Gelman A. \& Rubin D.B. (1992) Stat.Sc. {\bf 7}, 457.

\bibitem{H1} Hu W. \& Sugiyama N. (1995) PR D{\bf 51}, 2599.

\bibitem{K1} Kamenshchik A., Moschella U. \& Pasquier V.  (2001) PL B{\bf
511}, 265.

\bibitem{Ko} Komatsu Y. et al (2008) arXiv:0803.0547v1 (astro--ph)
and APJ Suppl (in press).

\bibitem{C2} La Vacca G. \& Colombo L.P.L. (2007) JCAP {\bf 0804}, 007
(2008).

\bibitem{C2a} La Vacca G., S.A. Bonometto \& Colombo
L.P.L., arXiv:0810.0127 (astro--ph) and New Astr. (submitted)

\bibitem{L1} Lewis A. \& Bridle S. (2002) PR D{\bf 66}, 103511.

\bibitem{M1} Mainini R. \& Bonometto S.A. (2007) JCAP {\bf 06}, 020.

\bibitem{M1b} Mainini R. \& Bonometto S.A. (2004)  PR Lett.~{\bf 93},
121301.

\bibitem{M2} Mainini R., Colombo L.P.L. \& Bonometto S.A. (2005) ApJ
{\bf 632}, 691.

\bibitem{M2a} Majerotto E., Sapone D. \& Amendola L. (2006)
astro-ph/0610806.

\bibitem{M3} Maccio' A.V., Quercellini C., Mainini R., Amendola L. \&
Bonometto S.A. (2004) PR D{\bf 69}, 123516.

\bibitem{P1} Perlmutter S. et al. (1997) ApJ {\bf 483}, 565;
(1998) Nature {\bf 391}, 51.

\bibitem{percival06} Percival  W.J. \& Brown M.L. (2006) MNRAS {\bf 372}, 1104.

\bibitem{R0} Ratra B. \& Peebles P.J.E. (1988) PR D{\bf 37}, 3406.

\bibitem{Re} Refregier et al. (2006) Procs. of SPIE symposium
"Astronomical Telescopes and Instrumentation", Orlando, May 2006,
astro-ph/0610062.

\bibitem{Re1} Refregier et al. (2008) arXiv:0807.4036 and to appear on
Procs. of SPIE symposium "Astronomical Telescopes and
Instrumentation", Marseille, June 2006


\bibitem{R1} Riess et al. (1997) ApJ {\bf 114}, 722.

\bibitem{S1} Spergel D.N. et al. (2007) ApJ Suppl. {\bf 170}, 377.

\bibitem{W1} Wetterich C. (1988) Nuc.Phys. B{\bf 302}, 483; (1995) A\&A
{\bf 301}, 32.

\end{thebibliography}
\end{document}